\documentclass[12pt,a4paper,twoside]{article}

\usepackage{fancyhdr,graphicx,subfigure}
\usepackage{amsmath,amsfonts,amsthm}
\usepackage[abs]{overpic}

\DeclareMathOperator{\trace}{Tr}
\DeclareMathOperator{\sgn}{sgn}

\DeclareMathOperator{\diag}{diag}

\newcommand\ie{\emph{i.e. }}

\newcommand{\abs}[1]{\left \lvert #1 \right \rvert}
\newcommand{\norm}[1]{\left  \lVert #1 \right \rVert}
\newcommand{\UN}{\mathrm{U}(N)}
\newcommand{\ON}{\mathrm{O}(N)}
\newcommand{\GL}{\mathrm{GL}(N,\mathbb{C})}
\newcommand{\SpN}{\mathrm{Sp}(N)}
\newcommand{\USpN}{\mathrm{USp}(2N)}
\newcommand{\TN}{\mathrm{T}(N)}
\newcommand{\GN}{\mathrm{\Gamma}(N)}
\newcommand{\LN}{\Lambda(N)}
\newcommand{\TNH}{\mathrm{T}(N,\mathbb{H})}
\newcommand{\GNH}{\mathrm{\Gamma}(N,\mathbb{H})}
\newcommand{\LNH}{\Lambda(N,\mathbb{H})}
\newcommand{\dmh}{d\mu_{\mathrm{H}}}
\newcommand{\dmg}{d\mu_{\mathrm{G}}}

\newtheorem{theorem}{Theorem}
\newtheorem{lemma}{Lemma}
\newtheorem{corollary}{Corollary}

\numberwithin{equation}{section}

\begin{document}

\title{How to generate random matrices from the classical compact
groups}
\author{Francesco Mezzadri}
\date{}
\maketitle

\begin{abstract}
  We discuss how to generate random unitary matrices from the
  classical compact groups $\UN$, $\ON$ and $\USpN$ with probability
  distributions given by the respective invariant measures. The
  algorithm is straightforward to implement using standard linear algebra
  packages. This approach extends to the Dyson circular ensembles too.
  This article is based on a lecture given by the author at the summer
  school on Number Theory and Random Matrix Theory held at the
  University of Rochester in June 2006.  The exposition is addressed
  to a general mathematical audience.
\end{abstract}

\section{Introduction}
\label{intro}

Since Random Matrix Theory (RMT) was introduced by
Wishart~\cite{Wis28} in 1928, it has found applications in a variety
of areas of physics, pure and applied mathematics, probability,
statistics and engineering. Few examples --- far from being exhaustive
--- include: analytic number theory, combinatorics, graph theory,
multivariate statistics, nuclear physics, quantum chaos, quantum
information, statistical mechanics, structural dynamics and wireless
telecommunications.  The main reasons for the ever growing success of
RMT are mainly two. Firstly, in the limit of large matrix dimension
the statistical correlations of the spectra of a family, or ensemble,
of matrices are independent of the probability distribution that
defines the ensemble, but depend only on the invariant properties of
such a distribution. As a consequence random matrices turn out to be
very accurate models for a large number of mathematical and physical
problems. Secondly, RMT techniques allow analytical computations to an
extent that is often impossible to achieve in the contexts that they
are modelling. This predictive ability of RMT is particularly powerful
whenever in the original problem there are not any natural parameters
to average over.

Although the advantage of using RMT lies in the possibility of
computing explicit mathematical and physical quantities
analytically, it is sometimes necessary to resort to numerical
simulations.  The purpose of this article is twofold. Firstly, we
provide the reader with a simple method for generating random
matrices from the classical compact groups that most
mathematicians --- not necessarily familiar with computer
programming~--- should be able to implement in a code of only few
lines. This is achieved in section~\ref{corr_alg}. Secondly, we
discuss in detail the main ideas, which turn out be fairly general
and quite interesting, behind this algorithm.

An $N \times N$ unitary matrix $U=(u_{jk})$ is defined by the relation
$U^*U=UU^*=I$, which in terms of the matrix elements reads
\begin{equation}
  \label{un_con}
    \sum_{k=1}^N u^*_{jk}u_{kl}  =\sum_{k=1}^N\overline u_{kj}u_{kl}
    = \delta_{jl} \quad \text{and}
    \quad \sum_{k=1}^N u_{jk}u^*_{kl} = \sum_{k=1}^N\overline u_{jk}u_{lk}=
    \delta_{jl},
\end{equation}
where $U^*=(u^*_{jk})$ is the conjugate transpose of $U$, \ie
$u^*_{jk}=\overline u_{kj}$. In this article we will use the
symbol~$\bar{}$~to denote complex conjugation, in order to
distinguish it from ${}^*$, which is reserved to the conjugate
transpose of a matrix. The constraints~\eqref{un_con} simply state
that the columns (rows) of a unitary matrix form an orthonormal
basis in $\mathbb{C}^N$. The set $\UN$ of unitary matrices forms a
compact Lie group whose real dimension is $N^2$; it is then made
into a probability space by assigning as a distribution the unique
measure invariant under group multiplication, known as \emph{Haar
measure.}  Such a probability space is often referred to as
Circular Unitary Ensemble (CUE).

Usually the correct ensemble to model a particular situation depends
on the symmetries of the problem. Ensembles of unitary matrices are
constructed in two steps: we first identify a subset $\mathsf{U}
\subset\UN$ by imposing further restrictions on $U$; then we assign to
$\mathsf{U}$ a probability measure with the appropriate invariant
properties. As well as $\UN$, we will discuss how to generate random
matrices from the orthogonal $\ON$ and unitary symplectic $\USpN$
groups with probability distributions given by the respective unique
invariant measures.  We shall also consider the two remaining
$\emph{Dyson circular ensembles}$~\cite{Dys62}, namely the Circular
Orthogonal Ensemble (COE) and Circular Symplectic Ensemble (CSE).
Other symmetric spaces appear in the applications~\cite{Zir96}, but we
will not concern ourselves with them.

%In recent years the classical compact groups $\UN$, $\ON$ and
%$\USpN$ have played a fundamental role in the applications of RMT
%to analytic number theory. Indeed, the introduction of RMT
%techniques in the the theory of the Riemann zeta function and
%other $L$-functions has produced a wealth of results that would
%have been inconceivable before. The the work of Katz and
%Sarnak~\cite{KS99}, Keating and Snaith~\cite{KS00a,KS00b} and
%Rubinstein~\cite{Rub01} has shown beyond doubt that the correct
%ensembles to model $L$-functions are the classical compact groups.
%(For a series of review articles on the subject see
%ref.~\cite{MS05}.) Instead, the Dyson circular ensembles appear
%mainly in applications to quantum mechanics, whenever the emphasis
%is on time evolution.

%Another popular class of RMT ensembles are the Gaussian ensembles.
%However, the matrices in these spaces are not unitary. As the name
%suggests, their probability measures are defined by assigning to
%the independent parameters of a matrix a normal distribution. For
%example, the Gaussian Unitary Ensemble (GUE) is made of
%\emph{Hermitian matrices}, which obey the relation $H=H^*$; the
%diagonal elements of $H$ (which are real) and the real and
%imaginary parts of the upper triangular entries are independent
%normal random variables. Therefore, to generate numerically such
%matrices is trivial.

Writing an algorithm to generate random unitary matrices which is both
correct and numerically stable presents some pitfalls.  The reason
is that the conditions~\eqref{un_con} imply that the matrix elements
are not independent and thus are statistically correlated. The main
ideas discussed in this article are centred around the QR
decomposition and go back to Wedderburn~\cite{Wed75},
Heiberger~\cite{Hei78} (corrected by Tanner and Thisted~\cite{TT82}),
Stewart~\cite{Ste80} and Diaconis and Shahshahani~\cite{DS87}.
However, the technical literature may be difficult to access for a
reader without a background in numerical analysis or statistics, while
the implementation of such techniques is elementary. Another method
discussed in the literature involves an explicit representation of the
matrix elements of $U$ in terms of $N^2$ independent parameters (Euler
angles)~\cite{ZK94}, but it does not seem to be equally efficient or
convenient.

%The article is structured as follows: in section~\ref{haar_m} we
%introduce the notions of Haar measure; section~\ref{QRfact}
%discusses a pitfall that can easily deceives somebody blindly
%using routines in standard linear algebra packages; in
%section~\ref{corr_alg} we give a simple recipe to generate random
%unitary matrices that any undergraduate student who has attended a
%course in linear algebra should be able to implement;
%section~\ref{symp_gr} presents the quaternion formalism for
%$\USpN$; in section~\ref{H_r} we introduce the Householder
%reflections; section~\ref{gtheo} provides a natural and more
%general group theoretical interpretation of the techniques
%discussed previously; section~\ref{circular} ends the paper with a
%brief mention to the Dyson circular ensembles.

\section{Some examples and motivations}
\label{examples}

Before discussing how to generate random matrices it is helpful to
give few examples that show how they appear in the applications.

In quantum mechanics all the information about an isolated physical
system at a given time $t_0$ is contained in a state vector $\psi_0$
belonging to a Hilbert space $\mathcal{H}$ --- in general infinite
dimensional. The time evolution of $\psi_0$, \ie its dynamics, is
determined by a unitary operator $U$.  In other words, at a time $t >
t_0$, $\psi_0$ has evolved into
\begin{equation}
  \psi = U \psi_0.
\end{equation}
The fact that $U$ is unitary guarantees that
$\norm{\psi}=\norm{\psi_0}=1$, which is required by the probabilistic
interpretation of quantum mechanics.

If the dynamics is complicated --- as in heavy nuclei or in
quantum systems whose classical limits are characterized by a
chaotic dynamics
--- writing down an explicit expression for $U$ may be hopeless.
Therefore, we can attempt to replace $U$ by a random operator and
check if the predictions that we obtain are consistent with the
empirical observations. It is also reasonable to simplify the problem
even further and replace $U$ by a random unitary matrix of finite, but
large, dimension.  Then the main question is: What are the matrix
space and the probability distribution that best model our system?

In physics the symmetries of a problem are often known a priori, even
if the details of the dynamics remain obscure.  Now, suppose that our
system is invariant under time reversal but does not have any other
symmetry. From general considerations (see Mehta~\cite{Meh04} p. 36)
we know that $U$ is always conjugate by a unitary transformation to a
symmetric matrix. Therefore, we can always choose $U$ so that
\begin{equation}
  U = U^t,
\end{equation}
where $U^t$ denotes the transpose of $U$.  Since there are not
other symmetries in the problem, this is the only constraint that
we can impose.  Therefore, the appropriate matrices that model
this physical system should be symmetric. Let us denote by
$\mathsf{O}$ the set of unitary symmeytric matrices. If $U \in
\mathsf{O}$ it can be proved (see Meta~\cite{Meh04} p. 499) that
it admits the representation
\begin{equation}
  \label{coe_rep}
   U = WW^t, \quad W \in \UN.
\end{equation}
This factorization is not unique. Let $\ON$ be the group of real
matrices $O$ such that $OO^t=O^tO=I$ and set $W'=WO$.  By
definition we have
\begin{equation}
  U = W'W'^{t}=WOO^tW^t = WW^t.
\end{equation}
This statement is true also in the opposite direction: if $WW^t
=W'W'^t$ there exists an $O\in \ON$ such that $W'= WO$. Therefore,
$\mathsf{O}$ is isomorphic to the left coset space of $\ON$ in $\UN$,
\ie
\begin{equation}
  \label{eq:coe_def}
  \mathsf{O} \cong \UN / \ON.
\end{equation}

Since a measure space with total mass equal to one is a probability
space, in what follows we shall use the two terminologies
interchangeably. An ensemble of random matrices is defined by a matrix
space and a probability measure on it. We have found the former; we
are left to identify the latter.  Haar measure, which will be
discussed in detail in section~\ref{haar_m}, provides a natural
probability distribution on $\UN$; `natural' in the sense that it
equally weighs different regions of $\UN$, thus it behaves like a
uniform distribution.  From the factorization~\eqref{coe_rep} the
probability distribution on $\UN$ induces a measure on $\mathsf{O}$.
As a consequence, if $W$ is Haar distributed the resulting measure on
$\mathsf{O}$ will be uniform too.  In section~\ref{gtheo} we shall see
that such a measure is the unique probability distribution induced by
Haar measure on $\mathsf{O}$.  Therefore, it provides a natural choice
to model a time reversal invariant quantum system.  The space
$\mathsf{O}$ together with this measure is the COE ensemble.

If a quantum system does not have any symmetry, then there are no
restriction to $\UN$ and the natural choice of probability
distribution is Haar measure.  This is the CUE ensemble. If the system
is invariant under time reversal and has a half-integer spin, then the
appropriate ensemble is the CSE. The matrix space of the CSE is the
subset $\mathsf{S} \subset \mathrm{U}(2N)$ whose elements admit the
representation
\begin{equation}
  \label{cse_fact}
  U= -WJW^tJ, \quad W \in \mathrm{U}(2N),
\end{equation}
where
\begin{equation}
  \label{eq:skew_mat}
  J= \begin{pmatrix} 0 & I_N  \\ - I_N & 0 \end{pmatrix}.
\end{equation}
From the factorization~\eqref{cse_fact} the probability
distribution on $\mathrm{U}(2N)$ induces a measure on
$\mathsf{S}$.  As previously, such a measure is fixed by assigning
Haar measure to $\mathrm{U}(2N)$.

The set $\mathsf{S}$ is isomorphic to a coset space too. The unitary
symplectic group $\USpN$ is the subgroup of $\mathrm{U}(2N)$ whose
elements obey the relation
\begin{equation}
  \label{eq:symp_gr_def}
   SJS^t = J.
\end{equation}
Therefore, the matrix $U$ in equation~\eqref{cse_fact} does not change
if we replace $W$ with $W'=WS$, where $S \in \USpN$.  Similarly, if $W$
and $W'$ are such that
\begin{equation}
  \label{eq:fact_eq}
  U = -WJW^tJ = -W'JW'^tJ, \quad W, W' \in \mathrm{U}(2N),
\end{equation}
then $W'W^{-1} \in \USpN$.  Therefore,
\begin{equation}
\label{spmatcse}
\mathsf{S} \cong \mathrm{U}(2N)/\USpN.
\end{equation}
The probability distribution of the CSE is the unique invariant
measure induced on the coset space~\eqref{spmatcse} by Haar measure on
$\mathrm{U}(2N)$.

From equations~\eqref{coe_rep} and~\eqref{cse_fact} all we need to
generate random matrices in the CUE, COE and CSE ensembles is an
algorithm whose output is Haar distributed unitary matrices.  The
rest of this article will concentrate on generating random
matrices from all three classical compact groups $\UN$, $\ON$ and
$\USpN$ with probability distributions given by the respective
Haar measures. These groups are not only functional to
constructing matrices in the COE and CSE, but are also important
ensembles in their own right. Indeed, the work of
Montgomery~\cite{Mon73}, Odlyzko~\cite{Odl89}, Katz and
Sarnak~\cite{KS99}, Keating and Snaith~\cite{KS00a,KS00b} and
Rubinstein~\cite{Rub01} has shown beyond doubt that the local
statistical properties of the the Riemann zeta function and other
\textit{L}-functions can be modelled by the characteristic
polynomials of Haar distributed random matrices.  Over the last
few years the predictive power of this approach has brought about
an impressive progress in analytic number theory that could not
have been achieved with traditional techniques. (See~\cite{MS05}
for a collection of review articles in the subject.)

\section{Haar measure and invariance}
\label{haar_m}

Since the algorithm the we shall discuss is essentially based on the
invariant properties of Haar measure, in this section we introduce the
main concepts that are needed to understand how it works. We,
nevertheless, begin with another ensemble: \emph{the Ginibre
ensemble.} Besides being a simpler illustration of the ideas we need,
generating a matrix in the Ginibre ensemble is the first step toward
producing a random unitary matrix.

The space of matrices for the Ginibre ensemble is
$\mathrm{GL}(N,\mathbb{C})$, the set of all the invertible $N \times
N$ complex matrices $Z =(z_{jk})$; the matrix elements are independent
identically distributed (\emph{i.i.d.}) standard normal complex random
variables.  In other words, the probability density function
(\textit{p.d.f.}) of $z_{jk}$ is
\begin{equation}
\label{pdf_sncrv}
p(z_{jk}) = \frac{1}{\pi} e^{-\left \lvert z_{jk} \right \rvert^2}.
\end{equation}
By definition the matrix entries are statistically independent,
therefore the joint probability density function (\textit{j.p.d.f.}) for
the matrix elements is
\begin{equation}
P(Z) = \frac{1}{\pi^{N^2}} \prod_{j,k=1}^N e^{-\abs{z_{jk}}^2} =
\frac{1}{\pi^{N^2}} \exp \left(-
\sum_{j,k=1}^N\abs{z_{jk}}^2\right) = \frac{1}{\pi^{N^2}}\exp
\left(-\trace Z^*Z \right).
\end{equation}
Since $P(Z)$ is a probability density, it is normalized to one,
\ie
\begin{equation}
\int_{\mathbb{C}^{N^2}} P(Z) \, dZ = 1,
\end{equation}
where $dZ = \prod_{j,k=1}^N dx_{jk}dy_{jk}$ and $z_{jk} = x_{jk} +
i y_{jk}$. The \textit{j.p.d.f.}  $P(Z)$ contains all the statistical
information on the Ginibre ensemble.

Since $\mathbb{C}^{N\times N} \cong \mathbb{C}^{N^2}$, we will use the
two notations according to what is more appropriate for the context.
Thus, we can write
\begin{equation}
  \label{G_m}
  d \mu_{\mathrm{G}} (Z) = P(Z)\,dZ
\end{equation}
and think of $\dmg$ as an infinitesimal volume or measure in
$\mathbb{C}^{N^2}$.
%More precisely, $d\mu_{\mathrm{G}}$ is a
%complex differential $(N^2,N^2)$-form on $\mathbb{C}^{N^2}$.
If $f: \mathbb{C}^{N \times N} \longrightarrow \mathbb{C}^{N
\times N}$, we say that $d\mu_{\mathrm{G}}$ is invariant under $f$
if
\begin{equation}
d\mu_{\mathrm{G}}\bigl(f(Z)\bigr)=d\mu_{\mathrm{G}}(Z).
\end{equation}

\begin{lemma}
  \label{glinv}
  The measure of the Ginibre ensemble is invariant under left and
  right multiplication of $Z$ by arbitrary unitary matrices, i.e.
  \begin{equation}
    \label{Grinv}
    \dmg(UZ)= \dmg(ZV)= \dmg(Z), \quad U,V \in \UN.
  \end{equation}
 \end{lemma}
\begin{proof}
 First we need to show that $P(UZ)=P(Z)$; then we must prove
that the Jacobian of the map
\begin{equation}
\label{mlig}
Z \mapsto UZ
\end{equation}
(seen as a transformation in $\mathbb{C}^{N^2}$) is one. Since by
definition $U^*U=I$, we have
\begin{equation}
P(UZ) = \frac{1}{\pi^{N^2}} \exp \left(-\trace Z^*U^*UZ \right)=
 \frac{1}{\pi^{N^2}}\exp \left(-\trace Z^*Z \right)= P(Z).
\end{equation}
Now, the map~\eqref{mlig} is isomorphic to
\begin{equation}
X = \underset{N \text{times}}{\underbrace{U \oplus \cdots \oplus
U}}.
\end{equation}
It follows immediately that $X$ is a $N^2 \times N^2$ unitary
matrix, therefore $\abs{\det X}=1$. The proof of right invariance
is identical.
\end{proof}

Because the elements of a unitary matrix are not independent,
writing an explicit formula for the infinitesimal volume element
of $\UN$ is more complicated than for the Ginibre ensemble. An $N
\times N$ unitary matrix contains $2N^2$ real numbers and the
constraints~\eqref{un_con} form a system of $N^2$ real equations.
Therefore, $\UN$ is isomorphic to a $N^2$-dimensional manifold
embedded in $\mathbb{R}^{2N^2}$. Such a manifold is compact and
has a natural group structure that comes from matrix
multiplication. Thus, an infinitesimal volume element on $\UN$
will have the form
\begin{equation}
\label{ex_ex}
d \mu (U) = m(\alpha_1,\ldots,\alpha_{N^2})
d\alpha_1\cdots d\alpha_{N^2},
\end{equation}
where $\alpha_1,\ldots,\alpha_{N^2}$ are local coordinates on
$\UN$. Every compact Lie group has a unique (up to an arbitrary
constant) left and right invariant measure, known as \emph{Haar
measure.}  In other words, if we denote Haar measure on $\UN$ by
$\dmh(U)$, we have
\begin{equation}
\dmh(VU) = \dmh(UW) = \dmh(U),\quad V,W \in \UN.
\end{equation}
Although an explicit expression for Haar measure on $\UN$ in terms of
local coordinates can be written down (see
\.{Z}yczkowski and Kus~\cite{ZK94} for a formula), we
will see that in order to generate matrices distributed with Haar
measure we only need to know that is invariant and unique.

Haar measure normalized to one is a natural choice for a probability
measure on a compact group because, being invariant under group
multiplication, any region of $\UN$ carries the same weight in a group
average. It is the analogue of the uniform density on a finite
interval. In order to understand this point consider the simplest
example: $\mathrm{U}(1)$.  It is the set $\{e^{i\theta}\}$ of the
complex numbers with modulo one, therefore it has the topology of the
unit circle $\mathbb{S}^1$.  Since in this case matrix multiplication
is simply addition $\bmod \: 2\pi$, $\mathrm{U}(1)$ is isomorphic to
the group of translations on $\mathbb{S}^1$.  A probability density
function that equally weighs any part of the unit circle is the
constant density $\rho(\theta)=1/(2\pi)$. This is the standard
Lebesgue measure, which is invariant under translations.  Therefore,
it is the unique Haar measure on $\mathrm{U}(1)$.
%\begin{figure}
%\label{translation}
%\centering
%\includegraphics[width=4 in]{translations.eps}
%\caption{An arc on the unit circle (chain curve) is translated by
%a fix amount.  Clearly, its length (Haar measure) is unchanged.}
%\end{figure}

Note that it is not possible to define an `unbiased' measure on a
non-compact manifold.  For example, we can provide a finite interval
with a constant \textit{p.d.f.} $\rho(x)$, but not the whole real line
$\mathbb{R}$, since the integral $\int_{-\infty}^{\infty}\rho(x)dx$
would diverge.

\section{The QR decomposition and a numerical experiment}
\label{QRfact}

By definition the columns of a $N \times N$ unitary matrix are
orthonormal vectors in $\mathbb{C}^N$.  Thus, if we take an arbitrary
complex $N\times N$ matrix $Z$ of full rank and apply the Gram-Schmidt
orthonormalization to its columns, the resulting matrix $Q$ is
unitary. It turns out that if the entries of $Z$ ares \emph{i.i.d.}
standard complex normal random variables, \ie if $Z$ belongs to the
Ginibre ensemble, then $Q$ is distributed with Haar measure (see
Eaton~\cite{Eat83}, p. 234, for a proof).  Unfortunately, the
implementation of this algorithm is numerically unstable. However, we
may observe that
\begin{equation}
Z = QR,
\end{equation}
where $R$ is upper-triangular and invertible.  In other words, the
Gram-Schmidt algorithm realizes the \emph{QR decomposition.} This
factorization is widely used in numerical analysis to solve linear
least squares problems and as first step of a particular
eigenvalue algorithm. Indeed, every linear algebra package has a
routine that implements it. In most cases, however, the algorithm
adopted is not the Gram-Schmidt orthonormalization but uses the
\emph{Householder reflections}, which are numerically stable.

Because of this simple observation, at first one might be tempted to
produce a matrix in the Ginibre ensemble and then to use a black box
QR decomposition routine. Writing such a code is straightforward.  For
example, if we choose the SciPy library in Python, we may implement
the following function:
\begin{tabbing}
\hspace{1cm} \= \texttt{from} \= \texttt{scipy import *} \\
\> \texttt{def wrong}\verb1_1\texttt{distribution(n):}\\
\>\> \texttt{''''''A Random matrix  with the wrong distribution''''''}\\
\>\>   \texttt{z = (randn(n,n) + 1j*randn(n,n))/sqrt(2.0)}\\
\>\>   \texttt{q,r = linalg.qr(z)}\\
\>\>    \texttt{return q}\\
\end{tabbing}
Unfortunately, the output is not distributed with Haar measure,
as it was observed by Edelman and Rao~\cite{ER05}. It is
instructive to give an explicit example of this phenomenon.

A unitary matrix can always be diagonalized in $\UN$.  Therefore,
its eigenvalues $\{e^{i\theta_1},\ldots,e^{i\theta_N}\}$ lie on
the unit circle. A classical calculation in RMT (see
Mehta~\cite{Meh04} pp.~203--205) consists of computing the
statistical correlations among the arguments $\theta_j$. The
simplest correlation function to determine is the density of the
eigenvalues $\rho(\theta)$, or --- as sometimes it is called
--- the one-point correlation. Since Haar measure is the analogue of a
uniform distribution, each set of eigenvalues must have the same
weight, therefore the normalized eigenvalue density is
\begin{equation}
\label{eigden}
\rho(\theta)= \frac{1}{2\pi}.
\end{equation}
It is important to point out that equation~\eqref{eigden} does not
mean that the eigenvalues are statistically uncorrelated.

Testing~\eqref{eigden} numerically is very simple. We generated
$10,000$ random unitary matrices using
\texttt{wrong}\verb1_1\texttt{distribution(n)}. The density of the
eigenvalues of such matrices is clearly not constant
(figure~1(a)). Figure~1(b) shows the histogram of the spacing
distribution, which deviates from the theoretical prediction too.
This statistics is often plotted because it encodes the knowledge of
all the spectral correlations and is easy to determine empirically.
For unitary matrices it is defined as follows.  Take the arguments of
the eigenvalues and order them in ascending order:
\begin{equation}
\theta_1 \le \theta_2 \le  \ldots \le \theta_N.
\end{equation}
The normalised distances, or spacings, between consecutive
eigenvalues are
\begin{equation}
s_j = \frac{N}{2\pi}(\theta_{j + 1} - \theta_j), \quad j=1,\ldots,
N.
\end{equation}
The spacing distribution $p(s)$ is the probability density of $s$.
(For a discussion on the spacing distribution see
Mehta~\cite{Meh04} p.~118.)

It is worth emphasising the QR decomposition is a standard
routine.  The most commonly known mathematical software packages
like Matlab, Mathematica, Maple, SciPy for Python essentially use
a combination of algorithms found in LAPACK routines.  Changing
software would not alter the outcome of this numerical experiment.

\begin{figure}
    \centering
    \label{fig2a}
        \subfigure[Eigenvalue density]{
    \begin{overpic}[scale=.75,unit=1mm,width=2.75in]{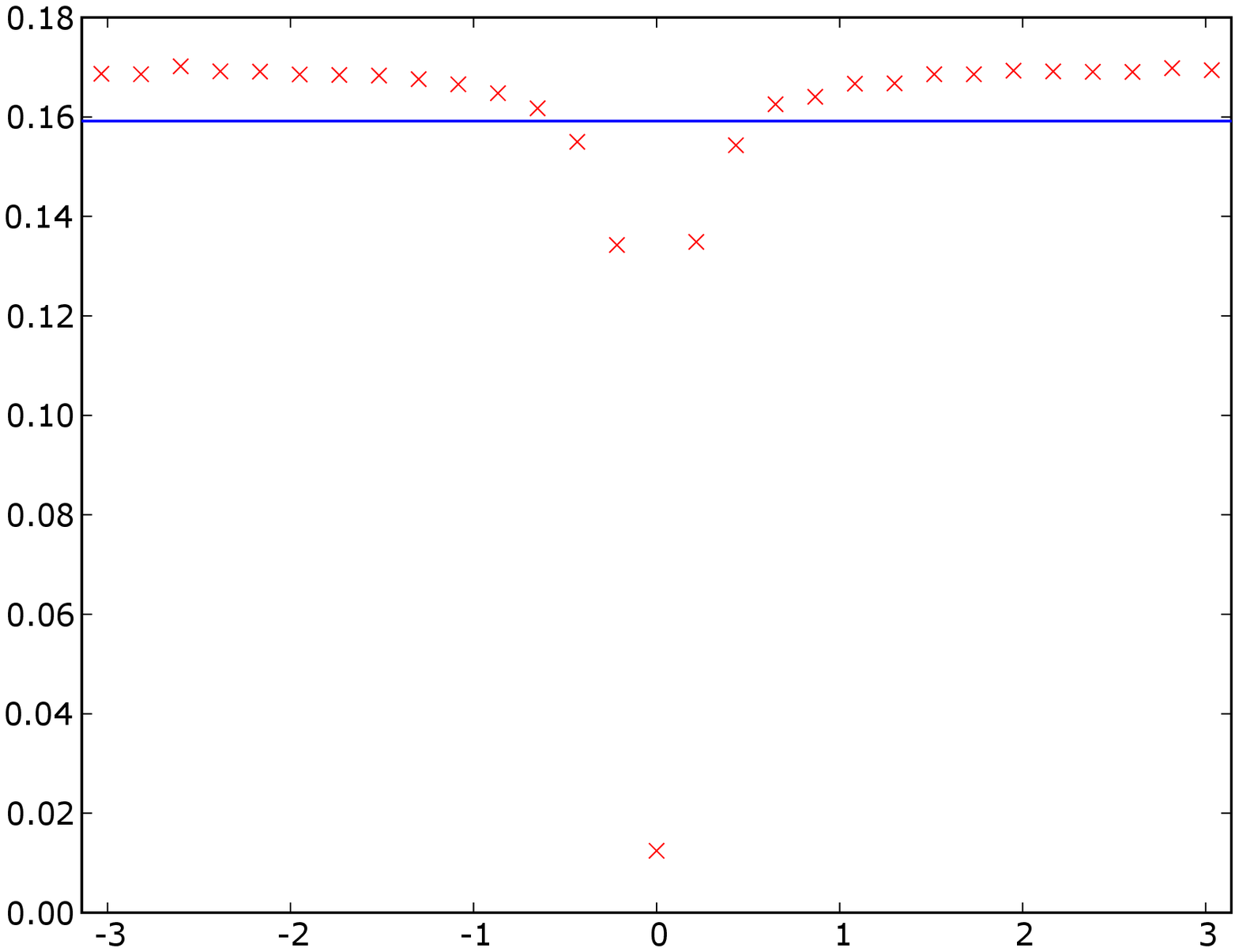}
        \put(35.5,1){{\tiny $\theta$}}
        \put(-1,26){{\tiny $\rho(\theta)$}}
    \end{overpic}}
    \hspace{.2in}
    \subfigure[Spacing distribution]{
    \begin{overpic}[scale=.75,unit=1mm,width=2.75in]{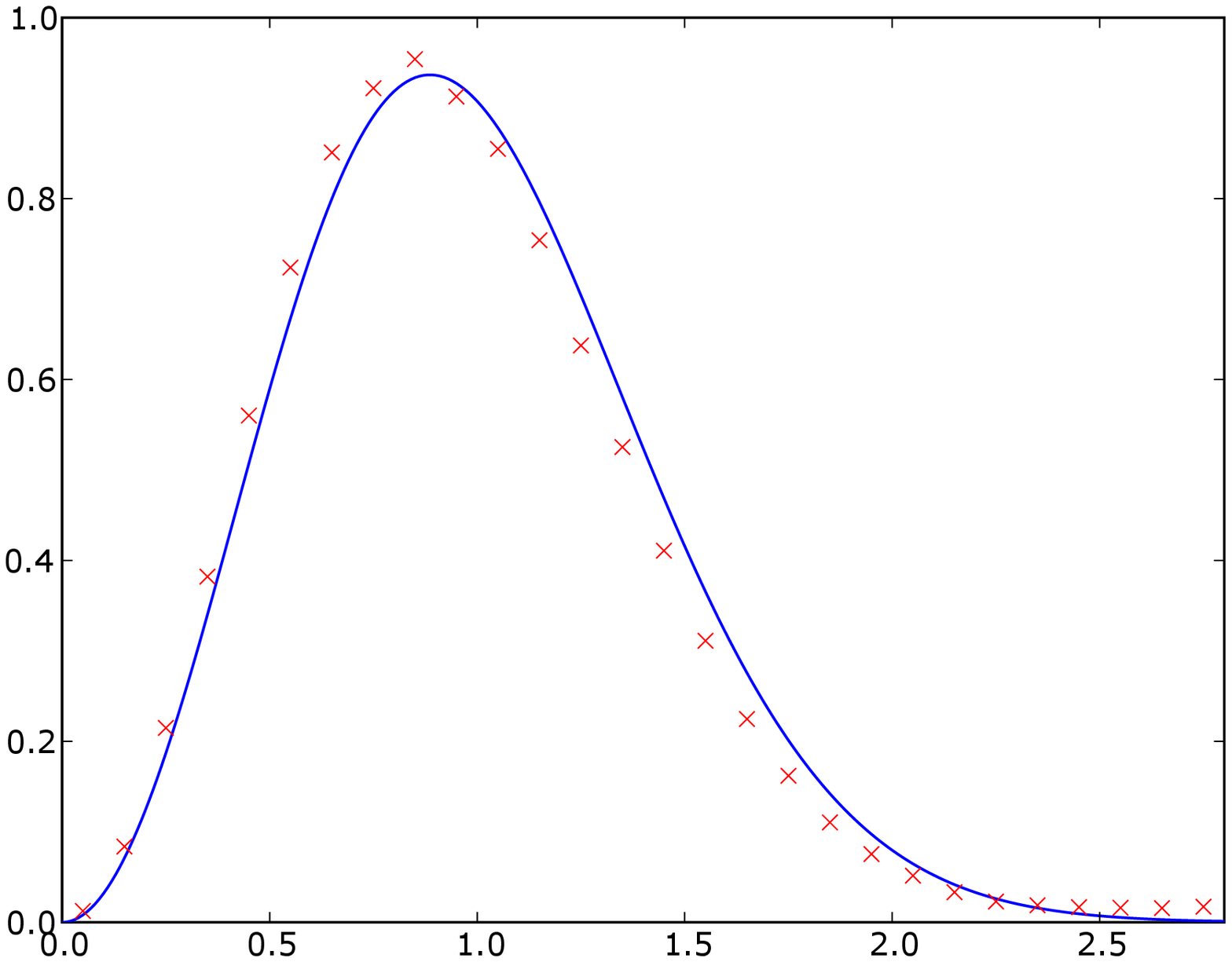}
        \put(36,1){{\tiny s}}
        \put(0,26){{\tiny $p(s)$}}
    \end{overpic}}
    \caption{Empirical histograms of the density of the eigenvalues
    and of the spacing distributions compared with the theoretical curves
     for the CUE. The data are computed from the eigenvalues of
     ten thousand $50 \times 50$ random unitary matrices obtained from the
    routine {\tt wrong\_distribution(n).}}.
\end{figure}

\section{A correct and efficient algorithm}
\label{corr_alg}

What is wrong with standard QR factorization routines? Where do
they differ from the Gram-Schmidt orthonormalization? Why is the
probability distribution of the output matrix not Haar measure?

The main problem is that QR decomposition is not unique.  Indeed, let
$Z \in \GL$ and suppose that $Z=QR$, where $Q$ is unitary and $R$ is
invertible and upper-triangular.  If
\begin{equation}
\label{un_diag}
\Lambda = \begin{pmatrix} e^{i\theta_1} & & \\ & \ddots & \\
& & e^{i\theta_N} \end{pmatrix} = \diag
\left(e^{i\theta_1},\ldots,e^{i\theta_N} \right),
\end{equation}
then
\begin{equation}
  \label{eq:relqr}
   Q' = Q\Lambda \quad \text{and} \quad R'=\Lambda^{-1}R
\end{equation}
are still unitary and upper-triangular respectively.  Furthermore,
\begin{equation}
\label{mult_valued}
Z = QR=Q'R'.
\end{equation}
Therefore, the QR decomposition defines a multi-valued map
\begin{equation}
\label{QRdecmap} \mathrm{QR}: \mathrm{GL}(N,\mathbb{C})
\longrightarrow \UN \times \TN,
\end{equation}
where $\TN$ denotes the group of invertible upper-triangular matrices.

In order to make the mapping~\eqref{QRdecmap} single-valued, we
need to specify the algorithm that achieves the factorization. In
most applications such a choice is dictated only by the
performance and stability of the code.  For our purposes, however,
the subset of \newline $\UN \times \TN$, in which the output of
the QR decomposition is chosen, is fundamental and we need to pay
particular attention to it. It is convenient from a mathematical
point of view to introduce a variation of the
mapping~\eqref{QRdecmap}, which is not only single-valued but also
one-to-one. In this way we will not have to refer all the times to
a specific algorithm. Indeed, the idea is that we should be able
to alter the output of a QR decomposition routine without even
knowing the algorithm implemented.

We first need
\begin{lemma}
Equation~\eqref{mult_valued} implies~\eqref{eq:relqr},
where $\Lambda \in \LN$ and $\Lambda(N)$ denotes the group of all
unitary diagonal matrices~\eqref{un_diag}.
\end{lemma}
\begin{proof}
Equation~\eqref{mult_valued} can be rearranged as
\begin{equation}
\label{intersec}
Q^{-1}Q' =R R'^{-1}.
\end{equation}
Since $\UN$ and $\TN$ are groups, both sides of
equations~\eqref{intersec} must belong to $\UN \cap \TN$. By
definition the inverse of a unitary matrix $U$ is its conjugate
transpose and the inverse of an upper-triangular matrix is
upper-triangular.  Therefore, if a matrix is both unitary and
upper-triangular it must be diagonal, \ie $\LN = \UN \cap \TN$.
\end{proof}

This lemma suggests that, more naturally, instead of the QR
factorization~\eqref{QRdecmap} we should consider a one-to-one map
\begin{equation}
\label{QRdecmap2} \overline{\mathrm{QR}}:
\mathrm{GL}(N,\mathbb{C}) \longrightarrow \UN \times \GN,
\end{equation}
where $\GN=\TN/\Lambda(N)$ is the right coset space of
$\Lambda(N)$ in $\TN$.  We construct~\eqref{QRdecmap2} as follows:
we first define it on a class of representatives of $\GN$ using
the QR factorization; then we extend it to the whole $\GN$.
However, since the QR decomposition is not unique, there is a
certain degree of arbitrariness in this definition. We need to
find a map under which the measure of the Ginibre ensemble induces
Haar measure on $\UN$. The main tool to achieve this goal is the
invariance under group multiplication of Haar measure and its
uniqueness. Thus, our choice of the
decomposition~\eqref{QRdecmap2} must be such that if
\begin{equation}
\label{ginv}
Z \mapsto (Q,\gamma) \quad \text{then} \quad UZ
\mapsto (UQ,\gamma)
\end{equation}
with the \emph{same} $\gamma \in \Gamma(N)$ and for \emph{any} $U
\in \UN$.  This property implies that left multiplication of $Z$
by a unitary matrix reduces, after the decomposition, to the left
action of $\UN$ into itself.  But lemma~\ref{glinv} states that
\begin{equation}
  \label{eq:gin_inv}
  \dmg(UZ) = \dmg(Z)
\end{equation}
for any $U\in \UN$.  As a consequence, if the
map~\eqref{QRdecmap2} satisfies~\eqref{ginv} the induced measure
on $\UN$ will be invariant under left multiplication too and
therefore must be Haar measure.

How do we construct the map~\eqref{QRdecmap2}? A class of
representatives of $\Gamma(N)$ can be chosen by fixing the arguments
of the elements of the main diagonal of $R \in \TN$. Let us impose
that such elements all be real and strictly positive.
Using~\eqref{eq:relqr} we can uniquely factorize any $Z \in \GL$ so
that the main diagonal of $R$ has this property.  It follows that if
$Z=QR$, then
\begin{equation}
  \label{eq:fact_rep}
   UZ = UQR, \quad U \in \UN.
\end{equation}
This QR decomposition of $UZ$ is unique within the chosen class of
representatives of $\GN$.  Therefore, the resulting
map~\eqref{QRdecmap2} obeys~\eqref{ginv}. Finally, we arrive at
\begin{theorem}
\label{salv-th} Suppose that the map~\eqref{QRdecmap2} satisfies
the hypothesis~\eqref{ginv}. Then, it decomposes the
measure~\eqref{G_m} of the Ginibre ensemble as
\begin{equation}
\label{propfac} \dmg(Z)=\dmh(Q)\times d \mu_{\GN}(\gamma).
\end{equation}
\end{theorem}
\begin{proof}
%The proof that the measure of the Ginibre ensemble decomposes as
%in equation~\eqref{propfac} is done in two steps: firstly, we
%prove it for a particular choice of a complete class of
%representatives of $\Gamma(N)$; secondly, we show that if we
%change the set of representatives the two measures remain the
%same. The measure $d\mu_{\Gamma(N)}$ can be explicitly determined
%(see Stewart~\cite{Ste80}).  However, since our concerns are
%limited only to Haar measure on $\UN $ we will not compute
%$d\mu_{\Gamma(N)}$.
We have
\begin{subequations}
\begin{align}
\dmg(UZ) & = \dmg(Z) && \text{by lemma~\ref{glinv}}\\
        & =  d \mu(UQ,\gamma)  = d \mu (Q,\gamma) && \text{by
        equation~\eqref{ginv}} \\
        \label{uptrm2}
         & = \dmh(Q) \times d\mu_{\Gamma(N)}(\gamma) && \text{by the uniqueness of Haar
         measure.}
\end{align}
\end{subequations}
\end{proof}

The choice of the class of representatives that we made coincides
exactly with outcome of the Gram-Schmidt orthonormalization.  The
output of standard QR decomposition routines are such that if $Z
\mapsto (Q,R)$ then $UZ \mapsto (Q',R')$ with $Q' \neq UQ$ and $R'
\neq R$. Therefore, the corresponding map~\eqref{QRdecmap2} does not
obey~\eqref{ginv} and theorem~\ref{salv-th} does not hold.

We can now give a recipe to create a random unitary matrix with
distribution given by Haar measure.
\begin{enumerate}
\item  Take an $N \times N$ complex matrix $Z$ whose entries are
complex standard normal random variables.
\item Feed $Z$ into \emph{any} QR decomposition routine.  Let
$(Q,R)$, where $Z = QR$, be the output.
\item  Create the following diagonal matrix
\begin{equation}
\Lambda= \begin{pmatrix} \frac{r_{11}}{\abs{r_{11}}} & & \\
                     & \ddots & \\
                     & & \frac{r_{NN}}{\abs{r_{NN}}}
  \end{pmatrix},
\end{equation}
where the $r_{jj}s$ are the diagonal elements of $R$.
\item The diagonal elements of $R' = \Lambda^{-1}R$ are
\emph{always} real and strictly positive, therefore the matrix $Q'
= Q\Lambda$ is distributed with Haar measure.
\end{enumerate}
The corresponding Python function is:
\begin{tabbing}
\hspace{1cm} \= \texttt{from} \= \texttt{scipy import *} \\
\> \texttt{def haar}\verb1_1\texttt{measure(n):}\\
\>\> \texttt{''''''A Random matrix distributed with Haar measure''''''}\\
\>\> \texttt{z = (randn(n,n) + 1j*randn(n,n))/sqrt(2.0)}\\
\>\> \texttt{q,r = linalg.qr(z)}\\
\>\> \texttt{d = diagonal(r)} \\
\>\> \texttt{ph = d/absolute(d)}\\
\>\> \texttt{q = multiply(q,ph,q)}\\
\>\>    \texttt{return q}
\end{tabbing}

If we repeat the numerical experiment discussed in
section~\ref{QRfact} using this routine, we obtain the histograms in
figure~2, which are consistent with the theoretical predictions.
\begin{figure}
    \centering
    \label{fig3a}
    \subfigure[Eigenvalue density]{
    \begin{overpic}[scale=.75,unit=1mm,width=2.75in]{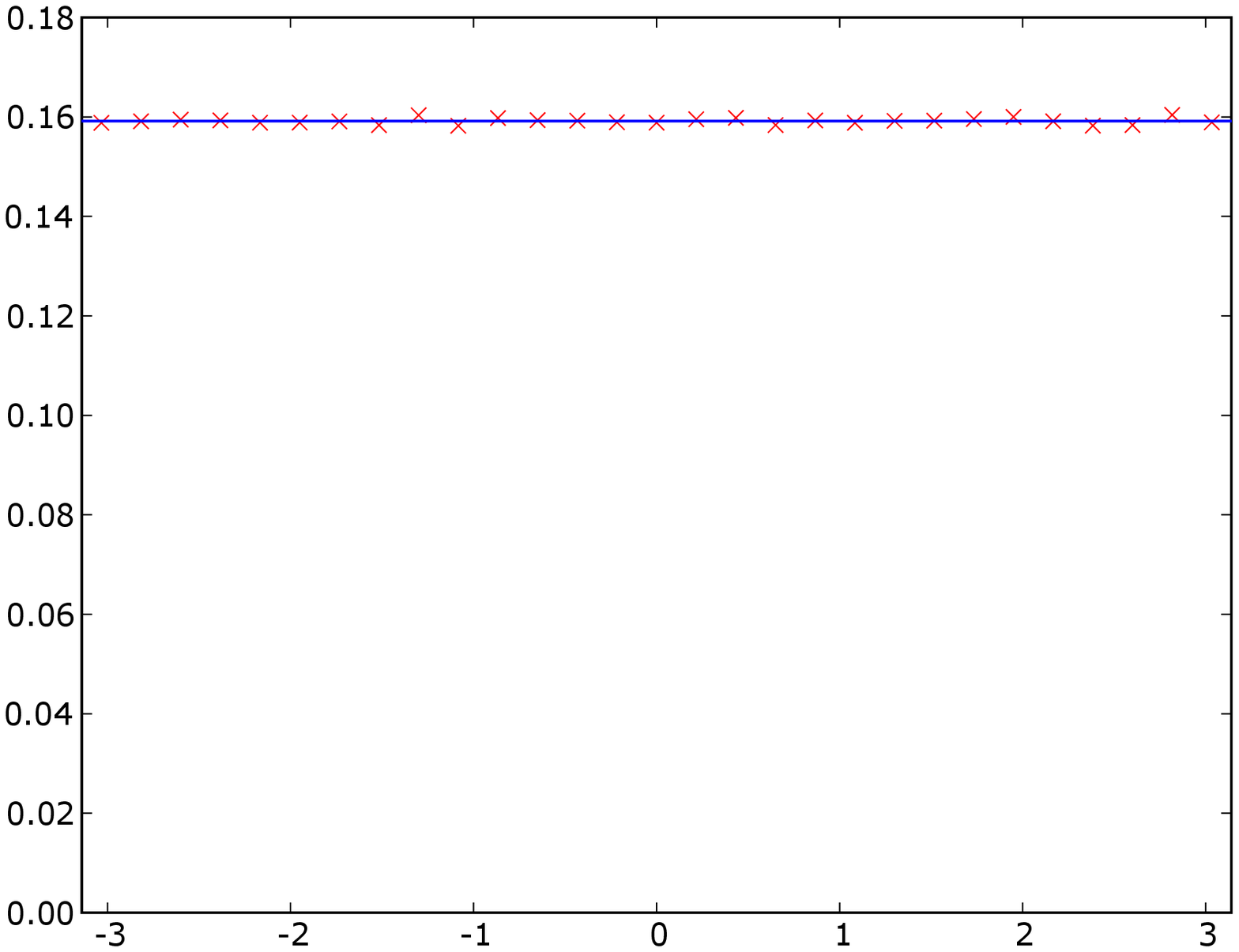}
        \put(35.5,1){{\tiny $\theta$}}
        \put(-1,26){{\tiny $\rho(\theta)$}}
    \end{overpic}}
    \hspace{.2in}
    \subfigure[Spacing distribution]{
    \begin{overpic}[scale=.75,unit=1mm,width=2.75in]{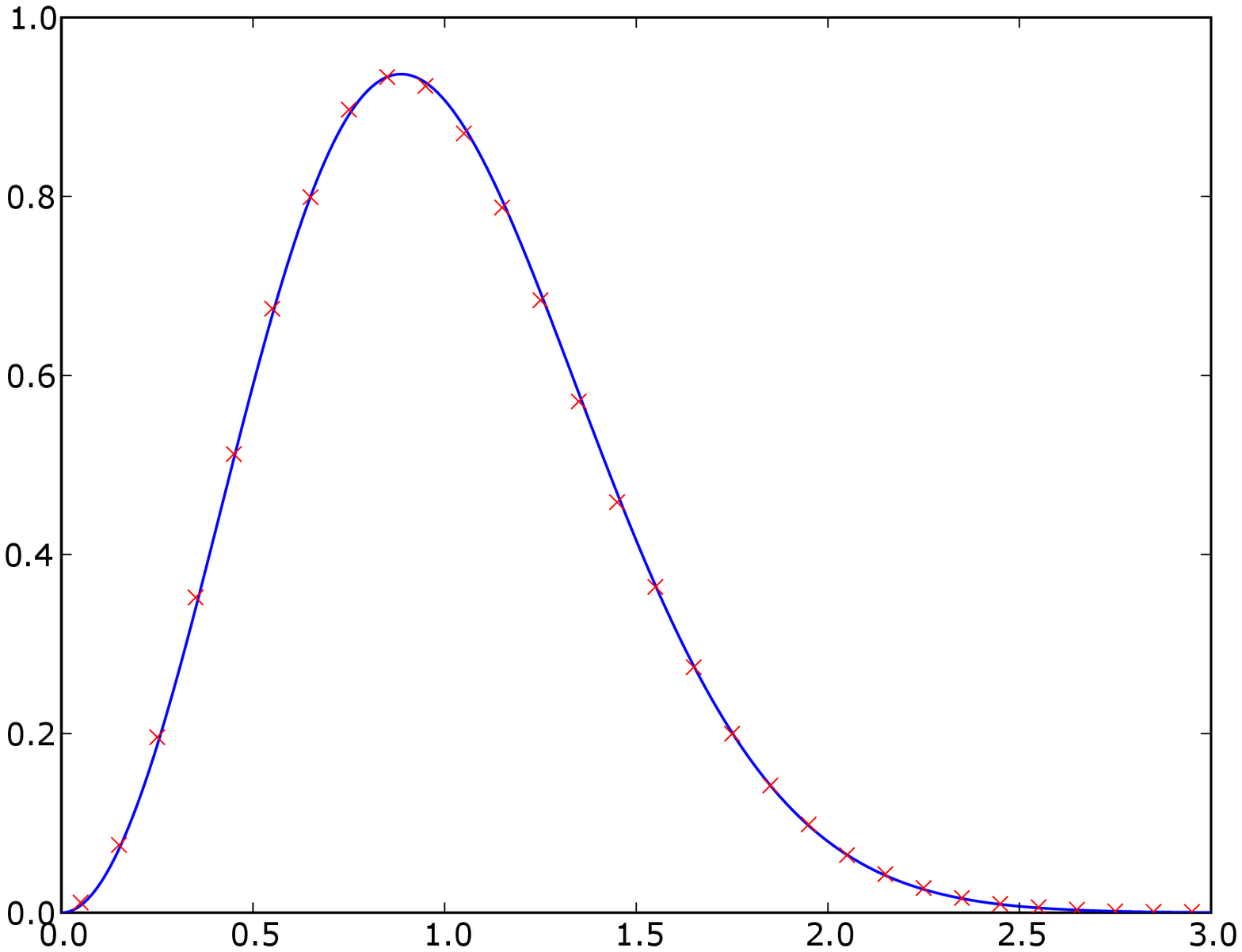}
        \put(36,1){{\tiny s}}
        \put(0,26){{\tiny $p(s)$}}
    \end{overpic}}
    \caption{Empirical histograms of the density of the eigenvalues
    and of the spacing distributions compared with the theoretical
    curves for the CUE.  The data are computed from the eigenvalues of
    ten thousand $50 \times 50$ random unitary matrices output of the
    function \texttt{haar\_measure(n)}.}
\end{figure}

\section{The unitary symplectic group $\USpN$}
\label{symp_gr}

Up to now we have only considered $\UN$.  The discussion for $\ON$ is
identical, except that the input matrix of the QR decomposition
routine must be real.  Unfortunately, however, for $\USpN$ there are
not any black box routines that we can use and we must put more effort
into writing an algorithm.

The algebra of unitary symplectic matrices can be rephrased in terms of
Hamilton's quaternions; it is convenient for our purposes to
use this formalism. A quaternion $q \in \mathbb{H}$ is a linear
combination
\begin{equation}
  \label{eq:quat_def}
  q = a \cdot 1 + bi_1 + ci_2 +di_3, \quad  a,b,c,d \in \mathbb{R},
\end{equation}
where $1$ is the identity and $i_1$, $i_2$, $i_3$ are the
quaternion units; they obey the algebra
\begin{equation}
  \label{eq:quat_alg}
   i_1^2 = i_2^2 = i_3^2 = i_1i_2i_3 = - 1.
\end{equation}
We can also define the conjugate of $q$,
\begin{equation}
    \label{qconj}
    \overline q = a \cdot 1 - bi_1 - ci_2  -di_3,
\end{equation}
as well the norm
\begin{equation}
  \label{eq:quat_norm}
  \norm{q}^2 = q\,\overline{q}= \overline{q}\,q = a^2 + b^2 + c^2 + d^2.
\end{equation}
When $c=d=0$, $\mathbb{H}$ reduces to $\mathbb{C}$ and $\overline q$
is simply the complex conjugate of $q$.

In analogy with $\mathbb{R}^N$ and $\mathbb{C}^N$ --- provided we are
careful with the fact that multiplication in $\mathbb{H}$ is not
commutative --- we can study the space $\mathbb{H}^N$. Elements in
$\mathbb{H}^N$ are $N$-tuples $\mathbf{q}=(q_1,\ldots,q_N)$.  The
bilinear map
\begin{equation}
  \label{qp}
  \left \langle \mathbf{p}, \mathbf{q} \right \rangle =
  \sum_{j=1}^N \overline p_j q_j, \quad \mathbf{p}, \mathbf{q} \in
  \mathbb{H}^N,
\end{equation}
is the analogue of the usual Hermitian inner product in
$\mathbb{C}^N$ and the norm of a quaternion vector is simply
\begin{equation}
  \norm{\mathbf{q}}^2 = \left \langle \mathbf{q},\mathbf{q}
  \right \rangle = \sum_{j=1}^N \norm{q_j}^2.
\end{equation}
Similarly, $\mathrm{GL}(N,\mathbb{H})$ is the group of all the
$N\times N$ invertible matrices with quaternion elements.

The quaternion units admit a representation in terms of the $2 \times
2$ matrices
\begin{subequations}
\begin{equation}
  \label{irrep}
  I_2 = \begin{pmatrix} 1 & 0 \\ 0 & 1 \end{pmatrix}, \quad
  e_1 = \begin{pmatrix} i & 0 \\ 0 & -i \end{pmatrix}, \quad
  e_2 = \begin{pmatrix} 0 & 1 \\ -1 & 0 \end{pmatrix} \quad
  \text{and} \quad
  e_3 = \begin{pmatrix} 0 & i \\ i & 0 \end{pmatrix},
\end{equation}
where
\begin{equation}
  \label{eq:unit_mapping}
  1 \mapsto I_2,\quad  i_1 \mapsto e_1, \quad i_2 \mapsto e_2 \quad
  \text{and} \quad i_3 \mapsto e_3.
\end{equation}
\end{subequations}
Thus, $q=a \cdot 1 + bi_1 + ci_2 + di_3$ is mapped into the
complex matrix
\begin{subequations}
  \label{elm}
  \begin{equation}
   \label{elmq}
    A= \begin{pmatrix} z & w \\ - \overline w & \overline z
   \end{pmatrix}
  \end{equation}
  where $z = a + ib$ and $w = c +id$.  In addition
  \begin{equation}
    \label{elmcq}
  \overline q \mapsto A^*= \begin{pmatrix} \overline z & -w \\
    \overline w & z \end{pmatrix} .
  \end{equation}
\end{subequations}
Equations~\eqref{elm} generalize to an arbitrary $N \times N$
quaternion matrix $\mathcal{Q}$, which can be represented in terms of
a $2N \times 2N$ complex matrix $Q$ using the decomposition
\begin{equation}
  \label{eq:quat_mat_rep}
  \mathcal{Q} \mapsto Q = Q_0 \otimes I_2 +  Q_1 \otimes e_1
  + Q_2 \otimes e_2 +  Q_3 \otimes e_3,
  \end{equation}
where $Q_0$, $Q_1$, $Q_2$ and $Q_3$ are arbitrary $N \times N$
\textit{real} matrices. Proceeding in the same fashion, if $\mathcal{Q} \in
\mathrm{GL}(N,\mathbb{H})$ we define its conjugate transpose
$\mathcal{Q}^*=(q^*_{jk})$ by setting $q^*_{jk}=
\overline{q}_{kj}$.

The symplectic group $\SpN$ is the subset of
$\mathrm{GL}(N,\mathbb{H})$ whose matrices satisfy the identity
$\mathcal{S}^*\mathcal{S}=\mathcal{S}\mathcal{S}^*=\mathcal{I}$.
%, which in terms
%of matrix elements reads
%\begin{subequations}
%  \label{eq:un_quat_con}
%  \begin{align}
%    \sum_{k=1}^N s^*_{jk}s_{kl} & =\sum_{k=1}^N\overline s_{kj}s_{kl}
%    = \delta_{jl},\\
%    \sum_{k=1}^N s_{jk}s^*_{kl} &= \sum_{k=1}^N\overline s_{jk}s_{lk}=
%    \delta_{jl},
%  \end{align}
%\end{subequations}
Because of the analogy between $\UN$ and $\SpN$, the latter is
sometimes called \textit{hyper-unitary group} and is denoted by
$\mathrm{U}(N,\mathbb{H})$.  The usefulness of the quaternion
algebra lies in
\begin{theorem}
\label{symp_is}
The groups $\SpN$ and $\USpN$ are isomorphic, i.e.
\begin{equation}
  \label{eq:is_un_spn}
  \USpN \cong \SpN.
\end{equation}
\end{theorem}
\begin{proof}
It is convenient to replace the skew-symmetric
matrix $J$ in the definition~\eqref{eq:symp_gr_def} with
\begin{equation}
  \Omega = \begin{pmatrix}
           0 & 1 & & & \\
           -1 & 0 & \ddots & & \\
           & \ddots & \ddots &  \ddots & \\
            & & \ddots & 0 & 1 \\
             & & & - 1 & 0 \\
            \end{pmatrix}
            = I \otimes e_2.
\end{equation}
This substitution is equivalent to a permutation of the rows and
columns of $J$, therefore it is simply a conjugation by a unitary
matrix.

We first prove that if $\mathcal{S} \in \SpN$, then its complex
representation $S$ belongs to $\USpN$. By
equation~\eqref{eq:quat_mat_rep} $\mathcal{S}^*$ is mapped to
\begin{equation}
  \label{eq:qut_trans}
   S_0^t \otimes I_2 -  S_1^t \otimes e_1 - S_2^t \otimes e_2
   -  S^t_3 \otimes e_3 = - \Omega \,S^t \,\Omega,
\end{equation}
which follows from the identities
\begin{equation}
  \label{eq:krock_prod_prop}
  (A \otimes B)^t = A^t \otimes B^t \quad \text{and}
  \quad (A \otimes B)(C \otimes D) = AC \otimes BD,
\end{equation}
and from the algebra~\eqref{eq:quat_alg} of the quaternion units.
As a consequence,
\begin{equation}
  \label{eq:symp_th}
  \mathcal{SS^*} \mapsto -S\,\Omega\,S^t \, \Omega = I.
\end{equation}
Therefore, the matrix $S$ is symplectic.  Combining
equations~\eqref{elmcq} and~\eqref{eq:qut_trans} gives
\begin{equation}
  \label{eq:un_sym_th1}
 -  \Omega\,S^t \, \Omega = S^*,
\end{equation}
thus $S \in \USpN$.

We now need to show that if $S \in \USpN$ then it is the
representation of a matrix $\mathcal{S} \in \SpN$. This statement
follows if we prove that $S$ admits a decomposition of the
form~\eqref{eq:quat_mat_rep}, where $S_0$, $S_1$, $S_2$ and $S_3$ $N
\times N$ must be \textit{real} matrices.  If this is true, then the
argument of the first part of the proof can simply be reversed.

Let us allow the coefficients $a$, $b$, $c$ and $d$ in the
definition~\eqref{eq:quat_def} to be complex numbers. The definitions
of conjugate quaternion~\eqref{qconj} and conjugate transpose of a
quaternion matrix, however, remain the same. The
matrices~\eqref{irrep} form a basis in $\mathbb{C}^{2 \times 2}$.
Therefore, any $2\times 2$ complex matrix can be represented as a
linear combination of $I_2$, $e_1$, $e_2$ and $e_3$.  Thus, any matrix
$Q \in \mathbb{C}^{2N \times 2N}$ admits a decomposition of the
form~\eqref{eq:quat_mat_rep}, but now the matrices $Q_0$, $Q_1$, $Q_2$
and $Q_3$ are allowed to be complex.  In other words, $Q$ is always
the representation of a quaternion matrix $\mathcal{Q}$, but in
general the quaternion units have complex coefficients.  The important
fact that we need to pay attention to is that
\begin{equation}
  \label{eq:conj_trans_st}
   \mathcal{Q}^* \mapsto Q^*,
\end{equation}
if and only if the coefficients of the quaternion units are real
numbers.  This is a straightforward consequence of the
representation~\eqref{elmq}.

Let $S \in \USpN$ be the complex representation of the quaternion
matrix $\mathcal{S}$, but assume that $\mathcal{S}^*$ is not mapped
into $S^*$.  It is still true, however, that
\begin{equation}
  \label{eq:trumap}
  \mathcal{S}^* \mapsto  -  \Omega\,S^t \, \Omega,
\end{equation}
because equation~\eqref{eq:qut_trans} is only a consequence of matrix
manipulations.  But since $S$ is unitary symplectic $S^* = -
\Omega\,S^t \, \Omega$, which is a contradiction.
\end{proof}

The algebra of $\SpN$ is the generalization to Hamilton's quaternions
of the algebra of $\UN$.  Therefore, it is not surprising that the
discussion of section~\ref{corr_alg} is not affected by replacing
$\mathrm{GL}(N,\mathbb{C})$ and $\UN$ with $\mathrm{GL}(N,\mathbb{H})$
and $\SpN$ respectively. Thus, since $\SpN$ and $\USpN$ are
isomorphic, $\USpN$ and $\SpN$ have the same Haar measure $\dmh$.  In
particular, we can introduce the quaternion Ginibre ensemble, which is
the set $\mathrm{GL}(N,\mathbb{H})$ equipped with the probability
density
\begin{equation}
  P(\mathcal{Z})=\frac{1}{\pi^{2N^2}}\exp \left(-\trace
    \mathcal{Z}^*\mathcal{Z} \right)=\frac{1}{\pi^{2N^2}} \exp \left(-
    \sum_{j,k=1}^N\norm{z_{jk}}^2\right).
\end{equation}

Quaternion matrices can be factorized by the QR decomposition too:
for any \newline $\mathcal{Z} \in \mathrm{GL}(N,\mathbb{H})$ we
can always write
\begin{equation}
  \mathcal{Z}= \mathcal{Q}\mathcal{R},
\end{equation}
where $\mathcal{Q} \in \SpN$ and $\mathcal{R}$ is an invertible and
upper-triangular quaternion matrix. Now, let
\begin{equation}
  \LNH = \Bigl \{ \Lambda \in \TNH \bigl \lvert \: \Lambda = \diag
  \left(q_1,\ldots,q_N\right), \quad  \norm{q_j}=1, \quad j
  =1,\ldots,N \Bigr \},
\end{equation}
where $\TNH$ is the group of invertible upper-triangular quaternion
matrices. Furthermore, let $\GNH=\TNH/\LNH$ be the right coset space
of $\Lambda(N,\mathbb{H})$ in $\TNH$. We have the following
\begin{theorem}
\label{qrfacth2}
There exists a one-to-one map
\begin{equation}
\label{quatdec} \mathcal{QR}: \mathrm{GL}(N,\mathbb{H})
\longrightarrow \SpN \times \Gamma(N,\mathbb{H})
\end{equation}
such that
\begin{equation}
    \label{imp_inv_propq}
    \mathcal{Z} \mapsto (\mathcal{Q},\gamma) \quad \text{and}
    \quad \mathcal{UZ} \mapsto (\mathcal{UQ},\gamma),
\end{equation}
where $\gamma = \LNH\mathcal{R}$.  Furthermore, it factorizes the
measure $\dmg$ of the Ginibre ensemble as
\begin{equation}
    \label{propfac2}
    \dmg(\mathcal{Z})=\dmh(\mathcal{Q}) \times d\mu_{\GNH}(\gamma).
\end{equation}
\end{theorem}

We leave proving these generalizations as an exercise for the
reader.

\section{Householder reflections}
\label{H_r}

Theorem~\ref{qrfacth2} provides us with the theoretical tools to
generate a random matrix in $\USpN$.  However, when we implement these
results in computer code, we need to devise an algorithm whose output
satisfies the condition~\eqref{imp_inv_propq}. The first one that
comes to one's mind is the Gram-Schmidt orthonormalization. But given
that black box routines for quaternion matrices do not exists on the
market, and that we are forced to write the complete code ourselves,
we may as well choose one that is numerically stable and which, as it
turns out, requires the same effort. The most common algorithm that
achieves the QR decomposition uses the \emph{Householder reflections.}
For the sake of clarity, we will discuss this method for $\ON$; the
generalizations to $\UN$ and $\SpN$ are straightforward.

Given an arbitrary vector $\mathbf{v} \in \mathbb{R}^m$, the main
idea of the Householder reflections is to construct a simple
orthogonal transformation $H_m$ (dependent on $\mathbf{v}$) such
that
\begin{equation}
  \label{Htdef}
  H_m \mathbf{v}= \norm{\mathbf{v}} \mathbf{e_1},
\end{equation}
where $\mathbf{e_1} = (1,0,\ldots,0) \in \mathbb{R}^m$.  For any real
matrix $X=(x_{jk})$, $H_N$ is determined by replacing $\mathbf{v}$ in
equation~\eqref{Htdef} with the first column of $X$.  The product
$H_NX$ will have the structure
\begin{equation}
  \label{prmt}
  H_NX = \begin{pmatrix} r_{11} & * & \cdots & * \\
    0  & * & \cdots & * \\
    \vdots & \vdots & & \vdots \\
    0 & *&  \cdots & *
      \end{pmatrix},
\end{equation}
where
\begin{equation}
r_{11}= \norm{\mathbf{v}} = \sqrt{\textstyle{\sum_{j=1}^N
x_{j1}^2}}.
\end{equation}
Then, define the matrix
\begin{equation}
\tilde H_{N-1}=\left( \begin{array}{cc}
    1 & 0 \\
    0 &  \begin{array}{|ccc|}
     \hline
     & & \\
     &  H_{N-1} & \\
     & & \\
     \hline
     \end{array}
\end{array}
\right),
\end{equation}
where
\begin{equation}
  H_{N-1}(\mathbf{v}')\mathbf{v}'=\norm{ \mathbf{v}'}\mathbf{e}_1.
\end{equation}
In this case $\mathbf{v}'$ is the $(N-1)$-dimensional vector
obtained by dropping the first element of the second column of the
matrix~\eqref{prmt}.  We proceed in this fashion until the matrix
\begin{equation}
  \label{uptri}
  R=\tilde H_1\tilde H_2\cdots \tilde H_{N-1} H_N X
\end{equation}
is upper-triangular with diagonal entries $r_{11}, \ldots,
r_{NN}$. The product
\begin{equation}
  \label{ortmtx}
  Q = H_N^t \tilde H_{N-1}^t \cdots \tilde H_2^t \tilde H_1^t
\end{equation}
is by construction an orthogonal matrix. In
equations~\eqref{uptri} and~\eqref{ortmtx} $\tilde H_m$ denotes
the block matrix
\begin{equation}
  \tilde H_m= \begin{pmatrix} I_{N-m} &  \\ & H_m \end{pmatrix},
\end{equation}
where $H_m$ is defined in equation~\eqref{Htdef}.

The matrix $H_m$ is constructed using elementary geometry.
Consider a  vector in the plane, $\mathbf{v} =(x_1,x_2)$, and
assume, for simplicity, that $x_1 > 0$.  Furthermore, let
$\mathbf{\hat{u}}$ denote the unit vector along the interior
bisector of the angle $\phi$ that $\mathbf{v}$ makes with the
$x_1$-axis, \ie
\begin{equation}
  \mathbf{\hat{u}} = \frac{\mathbf{v} +
    \norm{\mathbf{v}}\mathbf{e}_1}{\bigl \lVert \mathbf{v} +
      \norm{\mathbf{v}}\mathbf{e}_1 \bigr \rVert},
\end{equation}
where $\mathbf{e}_1$ is the unit vector along the $x_1$-axis. The
reflection of $\mathbf{v}$ along  the direction of
$\mathbf{\hat{u}}$ is $\norm{\mathbf{v}}\mathbf{e}_1$ (see
figure~3).
\begin{figure}
\centering
\begin{overpic}[scale=.75,unit=1mm]{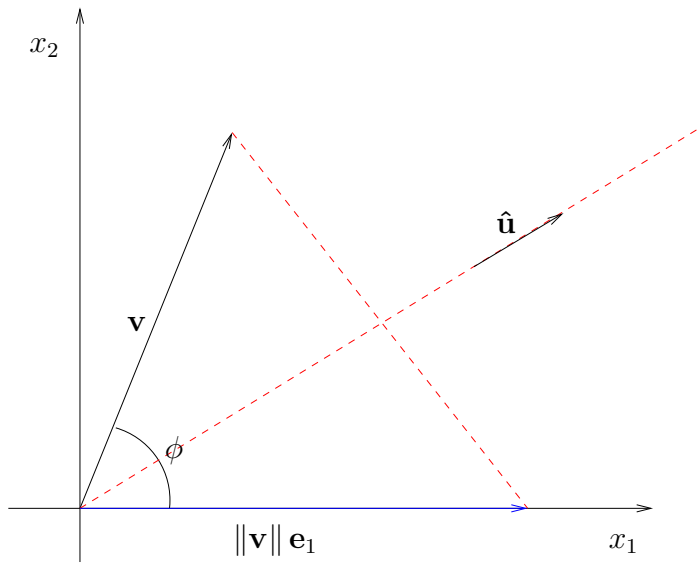}
   \put(80,2){$x_1$}
   \put(3,68){$x_2$}
   \put(21,14){$\phi$}
   \put(16,31){$\mathbf{v}$}
   \put(30,2){$\norm{\mathbf{v}}\mathbf{e}_1$}
   \put(65,44){$\mathbf{\hat u}$}
\end{overpic}
\caption{Householder reflection in $\mathbb{R}^2$.}
\end{figure}
Reflections are distance-preserving linear transformations,
therefore their representations in an orthonormal basis are
orthogonal matrices. In this simple example it can be constructed
from elementary linear algebra:
\begin{equation}
H_2(\mathbf{v}) = -I + 2\mathbf{\hat{u}}\mathbf{\hat{u}}^t.
\end{equation}
Finally, we obtain
\begin{equation}
\label{refpr} H_2(\mathbf{v}) \mathbf{v}=
\norm{\mathbf{v}}\mathbf{e}_1.
\end{equation}
It is worth noting that $H_2$ depends only on the direction of
$\mathbf{v}$ and not on its modulus. Thus we can rewrite
equation~\eqref{refpr} as
\begin{equation}
\label{refprb} H_2(\mathbf{\hat v}) \mathbf{\hat{v}}=
\mathbf{e}_1,
\end{equation}
where $\mathbf{\hat v} = \mathbf{v}/\norm{\mathbf{v}}$.

The generalization to arbitrary dimensions is straightforward. For
any vector $\mathbf{v}\in \mathbb{R}^m$, the Householder
reflection is defined as
\begin{equation}
  \label{hdg}
   H_m(\mathbf{\hat{v}})
   = \mp \left(I - 2\mathbf{\hat{u}}\mathbf{\hat{u}}^t\right),
\end{equation}
where
\begin{equation}
  \label{udef}
  \mathbf{\hat{u}} = \frac{\mathbf{v} \pm
    \norm{\mathbf{v}}\mathbf{e}_1}{\bigl \lVert \mathbf{v} \pm
      \norm{\mathbf{v}}\mathbf{e}_1 \bigr \rVert}.
\end{equation}
Furthermore, we have
\begin{equation}
  \label{refpr3}
  H_m(\mathbf{\hat{v}}) \mathbf{\hat{v}} = \mathbf{e}_1.
\end{equation}

How do we choose the sign in the right-hand side of
equation~\eqref{udef}? From a mathematical point of view such a
choice is irrelevant: in both cases $H_m(\mathbf{\hat v})$ maps
$\mathbf{v}$ into a vector whose only component different from
zero is the first one.  However, numerically it can be important.
The square of the denominator in~\eqref{udef} is
\begin{equation}
  \bigl \lVert \mathbf{v}
  \pm \norm{\mathbf{v}}\mathbf{e}_1 \bigr \rVert^2
  =2\norm{\mathbf{v}}\left(\norm{\mathbf{v}} \pm x_1\right),
\end{equation}
where $x_1$ is the first component of $\mathbf{v}$. If $x_1$ is
comparable in magnitude to $\norm{\mathbf{v}}$ and negative
(positive) and we choose the plus (minus) sign, then the term
\begin{equation}
  \label{eq:round-off}
  \norm{\mathbf{v}} \pm x_1,
\end{equation}
could very small and cancellations with significant round-off errors
may occur.  Therefore, the Householder transformation to be
implemented in computer code should be
\begin{equation}
\label{hdgn} H_m(\mathbf{\hat{v}}) = -\sgn(x_1) \left(I - 2
\mathbf{\hat{u}}\mathbf{\hat{u}}^t\right),
\end{equation}
where
\begin{equation}
\label{udefn} \mathbf{\hat{u}} = \frac{\mathbf{\hat v} +
\sgn(x_1) \mathbf{e}_1}{\norm{\mathbf{\hat v}  +
\sgn(x_1)\mathbf{e}_1}}.
\end{equation}

The additional factor of $\sgn(x_1)$ in the right-hand side of
equation~\eqref{hdgn} assures that there is no ambiguity in the
sign of the right-hand side of equation~\eqref{refpr3}.  In turn,
it guarantees that all the diagonal elements of the
upper-triangular matrix $R$ are positive. This is not the
definition of Householder reflection used in standard QR
decomposition routines.  Usually,
\begin{equation}
H'_m(\mathbf{\hat v}) = I - 2 \mathbf{\hat u}\mathbf{\hat u}^t,
\end{equation}
with the same $\mathbf{\hat u}$ as in~\eqref{udef}. Therefore,
\begin{equation}
H'_m(\mathbf{\hat v})\mathbf{\hat v} = \mp \mathbf{e}_1.
\end{equation}
As a consequence, the signs of the diagonal elements of $R$ are
random. This is the reason why the output of black box QR
decomposition routines must be modified in order to obtain
orthogonal matrices with the correct distribution.

Besides being numerically stable, this algorithm has another advantage
with respect to the Gram-Schmidt orthonormalization.  In most
applications of numerical analysis $Q$ need not be computed
explicitly, only $Q\mathbf{w}$ does, where $\mathbf{w}$ is a specific
vector.  Generating all the Householder reflections is an $O(N^2)$
process and computing $H_N \mathbf{w}$ requires $O(N)$ operations ---
it just evaluates the scalar product $(\mathbf{ \hat u},\mathbf{w})$.
Successively multiplying $H_{N}, \ldots, H_1$ into $\mathbf{w}$ is an
$O(N^2)$ process. Therefore, it takes in total $O(N^2)$ operations to
compute $Q\mathbf{w}$.  Instead, the Gram-Schmidt orthonormalization
is an $O(N^3)$ process.  However, if $Q$ is explicitly needed
computing the product~\eqref{ortmtx} requires $O(N^3)$ operations too.

The generalizations to $\UN$ and $\SpN$ are straightforward.  The
only differences are in the definitions of the Householder
reflections.  A suitable choice for $\UN$ is
\begin{equation}
  \label{htc}
  H_m(\mathbf{\hat{v}}) = - e^{-i\theta}\left(I-2
    \mathbf{\hat{u}}\mathbf{\hat{u}}^*\right).
\end{equation}
The unit vector $\mathbf{\hat{u}}$ is
\begin{equation}
  \label{cuv}
  \mathbf{\hat u} = \frac{\mathbf{\hat v} +
    e^{i\theta}\mathbf{e}_1}{\norm{\mathbf{\hat v} +
      e^{i\theta}\mathbf{e}_1}},
\end{equation}
where $\mathbf{v}=(x_1,\ldots,x_m)\in \mathbb{C}^m$ and
$x_1=e^{i\theta}\abs{x_1}$.  The matrix $H_m(\mathbf{\hat v})$ is
unitary and
\begin{equation}
  \label{impchn}
  H_m(\mathbf{\hat{v}}) \mathbf{\hat{v}} = \mathbf{e}_1.
\end{equation}
Note that the introduction of $e^{i\theta}$ in equations~\eqref{htc}
and~\eqref{cuv} takes into account both the potential cancellations
and the correct values of the arguments of the diagonal elements of
the upper-triangular matrix $R$: equation~\eqref{impchn} implies that
all the $r_{jj}$s are real and strictly positive.

For $\SpN$ we have
\begin{equation}
  \label{htq}
  H_m(\mathbf{\hat{v}}) = - \overline q \left(I-2
    \mathbf{\hat{u}}\mathbf{\hat{u}}^*\right).
\end{equation}
with
\begin{equation}
  \label{quv}
  \mathbf{\hat u} = \frac{\mathbf{\hat v} +
    q\mathbf{e}_1}{\norm{\mathbf{\hat v} + q\mathbf{e}_1}},
\end{equation}
where $\mathbf{v}=(x_1,\ldots,x_m) \in \mathbb{H}^m$ and
$x_1=q\norm{x_1}$. Also in this case
\begin{equation}
  \label{impch}
  H_m(\mathbf{\hat{v}}) \mathbf{\hat{v}} =
  \mathbf{e}_1.
\end{equation}

\section{A group theoretical interpretation}
\label{gtheo}

We now know how to generate random matrices from any of the
classical compact groups $\UN$, $\ON$ and $\SpN$.  In order to
achieve this goal, we have used little more than linear algebra.
However simple and convenient this approach is (after all linear
algebra plays a major role in writing most numerical algorithms),
it hides a natural group theoretical structure behind the
Householder reflections, which was uncovered by Diaconis and
Shahshahani~\cite{DS87}.  Indeed, generating a random matrix as a
product of Householder reflections is only one example of a more
general method that can be applied to any finite or compact Lie
group. Our purpose in this section is to give a flavour of this
perspective.  For the sake of clarity, as before, we will discuss the
orthogonal group $\ON$; the treatment of $\UN$ and $\SpN$
is, once again, almost identical.

The need of a more general and elegant approach arises also if one
observes that there is one feature of the QR decomposition that may
not be entirely satisfactory to a pure mathematician: Why in order to
generate a random point on a $N(N-1)/2$-dimensional manifold --- $\ON$
in this case --- do we need to generate $N^2$ random numbers?  It does
not look like the most efficient option, even if it is a luxury that
can be easily afforded on today's computers.

We will first show how the key ideas that we want to describe apply to
finite groups, as in this setting they are more transparent.  Suppose
that we need to generate a random element $g$ in a finite group
$\Gamma_N$. In this context, if $\Gamma_N$ has $p$ elements, uniform
distribution simply means that the probability of extracting any $g
\in \Gamma_N$ is $1/p$. In addition, we assume that there exists a
chain of subgroups of $\Gamma_N$:
\begin{equation}
  \Gamma_1 \subset \Gamma_2 \subset \cdots\subset \Gamma_N.
\end{equation}
In practical situations it is often easier to generate a random
element $\tilde g$ in a smaller subgroup, say $\Gamma_{m-1} \in
\Gamma_N$, than in $\Gamma_N$ itself; we may also know how to take a
random representative $g_m$ in the left coset $C_m =
\Gamma_m/\Gamma_{m-1}$. Now, write the decomposition
\begin{equation}
  \label{eq:sub_dec}
  \Gamma_m \cong C_m \times \Gamma_{m-1}.
\end{equation}
Once we have chosen a set of representatives of $C_m$, an element
$g \in \Gamma_m$ is uniquely factorized as $g = g_m \, \tilde g$,
where $g_m \in C_m$.  If both $g_m$ and $\tilde g$ are uniformly
distributed in $C_m$ and $\Gamma_{m-1}$ respectively, then $g$ is
uniformly distribute in  $\Gamma_m$.

We can apply this algorithm iteratively starting from $\Gamma_1$ and
eventually generate a random element in $\Gamma_N$.  In other words,
we are given the decomposition
\begin{equation}
  \label{eq:dec_in_subgr}
  \Gamma_N \cong C_{N} \times \cdots \times C_2 \times \Gamma_1.
\end{equation}
An element $g \in \Gamma_N$ has a unique representation as a product
\begin{equation}
  \label{eq:fact}
   g = g_N\cdots g_1,
\end{equation}
where $g_m$ is a representative in $C_m$.  If the $g_m$s are
uniformly distributed in $C_m$ so is $g$ in $\Gamma_N$.  This is known
as the \textit{subgroup algorithm}~\cite{DS87}.

This technique applies to random permutations of $N$ letters. The
chains of subgroups is
\begin{equation}
  \left \{\mathrm{Id}\right \}  \subset S_2 \subset  \cdots
  \subset S_N,
\end{equation}
where $S_m$ is the $m$-th symmetric group. Other examples include
generating random positions of Rubik's cube and random elements in
$\mathrm{GL}(N,\mathbb{F}_p)$, where $\mathbb{F}_p$ is a finite field
with $p$ elements.

For $\ON$ the decompositions~\eqref{eq:dec_in_subgr}
and~\eqref{eq:fact} are hidden behind the
factorization~\eqref{ortmtx} in terms of Householder reflections.
Indeed, the subgroup algorithm for $\ON$ is contained in
\begin{theorem}
  \label{hpr}
  Let $\mathbf{\hat v}_1,\ldots,\mathbf{\hat v}_N$ be uniformly
  distributed on $\mathbb{S}^{0},\ldots,\mathbb{S}^{N-1}$
  respectively, where
  \begin{equation}
    \label{eq:unit_sph}
    \mathbb{S}^{m-1}= \Bigl \{\mathbf{\hat v}_m = (x_1,\ldots,x_m)
    \in \mathbb{R}^m \bigl \lvert \:
    \textstyle{\sum_{j=1}^m x_j^2 = 1}\Bigr \}
  \end{equation}
  is the unit sphere in $\mathbb{R}^m$.  Furthermore, let
  $H_m(\mathbf{\hat v})$ be the $m$-th Householder reflection defined
  in equation~\eqref{hdgn}.  The product
  \begin{equation}
    \label{ortmatfac}
    O=H_N(\mathbf{\hat v}_N)H_{N-1}(\mathbf{\hat v}_{N-1})
    \cdots H_2(\mathbf{\hat v}_2)H_1(\mathbf{\hat v}_1)
  \end{equation}
  is a random orthogonal matrix with distribution given by Haar
  measure on $\ON$.
\end{theorem}
\begin{proof}
  Suppose we construct $O \in \ON$ distributed with Haar measure by
  factorizing a matrix $X$ in the Ginibre ensemble as described in
  section~\ref{H_r}.  The random matrix $O$ is the product of
  Householder reflections~\eqref{ortmtx} and each factor
  $H_m(\mathbf{\hat v}_m)$ is a function of the unit vector
  $\mathbf{\hat v}_m \in \mathbb{S}^{m-1}$ only. We need to show that
  such $\mathbf{\hat v}_m$s are independent and uniformly distributed
  in $\mathbb{S}^{m-1}$ for $m = 1,\ldots,N$.

  At each step in the construction of the upper-triangular
  matrix~\eqref{uptri}, the matrix multiplied by the $m$-th
  Householder reflection, \ie
  \begin{equation}
    \label{eq:int_fact}
     X_m= H_m(\mathbf{\hat v}_m) \cdots
     H_{N-1}(\mathbf{\hat v}_{N-1})H_N(\mathbf{\hat v}_N)X,
  \end{equation}
  is still in the Ginibre ensemble.  All its elements are, therefore,
  \textit{i.i.d.} normal random variables.  This is a consequence of
  the invariance
  \begin{equation}
    \dmg(OX)=\dmg(X), \quad O \in \ON,
  \end{equation}
  of the measure of the Ginibre ensemble.  Now, $\mathbf{\hat v}_m =
  (x_1,\ldots,x_m)$ is constructed by taking the $m$-th dimensional
  vector $\mathbf{v}_m$ obtained by dropping the first $N-m$ elements
  of the $(N - m + 1)$-th column of $X_m$. The components of
  $\mathbf{v}_m$ are \textit{i.i.d.} normal random variables.  It
  follows that the p.d.f. of $\mathbf{v}_m$ is
  \begin{equation}
    \label{eq:jpdfv}
    P(\mathbf{v}_m) = \frac{1}{\pi^{m/2}} 
    \prod_{j=1}^m \exp\left(-x_j^2 \right)
    = \frac{1}{\pi^{m/2}}\exp\left(-\sum_{j=1}^m x_j^2 \right)
    = \frac{1}{\pi^{m/2}}\exp\bigl(-\norm{\mathbf{v}_m}^2 \bigr).
  \end{equation}
  Since $P(\mathbf{v}_m)$ depends only on the length of
  $\mathbf{v}_m$, and not on any angular variable, the unit vector
  $\mathbf{\hat v}_m = \mathbf{v}_m/\norm{\mathbf{v}_m}$ is uniformly
  distributed in $\mathbb{S}^{m-1}$, and is statistically independent
  of $\mathbf{\hat v}_k$ for $k \neq m$.
\end{proof}
Theorem~\ref{hpr} is more transparent than relying on the QR
decomposition, which seems only a clever technical trick. If nothing
else, the counting of the number of degrees of freedom matches. In
fact, the dimension of the unit sphere $\mathbb{S}^{m-1}$ is $m-1$.
Thus, the total number of independent real parameters is
\begin{equation}
  \sum_{m=1}^{N} (m-1) = \frac{N(N-1)}{2}.
\end{equation}

Why is theorem~\ref{hpr} the subgroup algorithm for $\ON$? As we
shall see in theorem~\ref{quotient}, the
factorization~\eqref{ortmatfac} is unique --- provided that we
restrict to the definition~\eqref{hdgn} of the Householder
reflections.  This means that
\begin{equation}
\label{coset_dec}
\ON \cong \mathbb{S}^{N-1} \times \cdots \times
\mathbb{S}^1 \times \mathrm{O}(1),
\end{equation}
where
\begin{equation}
 \mathrm{O}(1) \cong \mathbb{S}^0  = \{-1,1\}.
\end{equation}
If we proceed by induction, we obtain
\begin{equation}
  \label{eq:or_fac2}
  \ON = \mathbb{S}^{N-1} \times \mathrm{O}(N-1).
\end{equation}
Therefore, a matrix $O\in \ON$ admits a unique representation as
\begin{equation}
\label{facmat} O = H_N(\mathbf{\hat v}_N) \Omega,
\end{equation}
where
\begin{equation}
  \label{eq:mat_omega}
  \Omega = \left( \begin{array}{cc}
      1 & 0 \\
      0 &  \begin{array}{|ccc|}
        \hline
        & & \\
        &  \tilde O & \\
        & & \\
        \hline
     \end{array}
     \end{array}\right)
\end{equation}
and $\tilde O \in \mathrm{O}(N-1)$. A consequence of
theorem~\ref{hpr} is that if $\mathbf{\hat v}_N$ is uniformly
distributed in $\mathbb{S}^{N-1}$ and $\tilde O$ is distributed
with Haar measure on $\mathrm{O}(N-1)$, then $O$ is Haar
distributed too. The final link with the subgroup algorithm is
given by
\begin{theorem}
  \label{quotient}
  The left coset space of $\mathrm{O}(N-1)$ in $\mathrm{O}(N)$ is
  isomorphic to $\mathbb{S}^{N-1}$, i.e.
  \begin{equation}
    \ON / \mathrm{O}(N-1) \cong \mathbb{S}^{N-1}.
  \end{equation}
  A complete class of representatives is provided by the
  map~\footnote{The Householder reflections defined in
    equation~\eqref{hdgn} are not continuous at $\mathbf{e}_1$.
    Indeed, it can be proved that there is no continuous choice of
    coset representatives. In section~\ref{H_r} this distinction was
    superfluous: if $\mathbf{v}$ is randomly generated, the
    probability that $\mathbf{v} = \alpha \mathbf{e}_1$ is zero.}
  $H_N: \mathbb{S}^{N-1} \longrightarrow \ON$,
  \begin{equation}
    \label{H_rm}
    H_N(\mathbf{\hat v})=
    \begin{cases} -\sgn(x_1)
      \left(I - 2 \mathbf{\hat{u}}\mathbf{\hat{u}}^t\right) &
      \text{\emph{if} $\mathbf{\hat v} \neq \mathbf{e}_1$,}\\
      I_N & \text{\emph{if} $\mathbf{\hat v}= \mathbf{e}_1$,}
    \end{cases}
  \end{equation}
  where
  \begin{equation}
    \mathbf{\hat{u}} = \frac{\mathbf{\hat v} + \sgn(x_1)
      \mathbf{e}_1}{\norm{\mathbf{\hat v}  + \sgn(x_1)\mathbf{e}_1}}
  \end{equation}
  and $x_1$ is the first component of $\mathbf{\hat v}$.
\end{theorem}
\begin{proof}
  The group of $N \times N$ matrices $\Omega$ defined in
  equation~\eqref{eq:mat_omega} is isomorphic to $\mathrm{O}(N-1)$.
  Since
  \begin{equation}
    \label{eq:omega_fix}
    \Omega \mathbf{e}_1 = \mathbf{e}_1,
  \end{equation}
 $\mathrm{O}(N-1)$ can be identified with the subgroup of
  $\ON$ that leave $\mathbf{e}_1$ invariant, \ie
\begin{equation}
  \mathrm{O}(N-1)= \bigl \{ O \in \ON \bigl \lvert \:
  O\mathbf{e}_1=\mathbf{e}_1 \bigr \}.
\end{equation}
Now, if two matrices $O$ and $O'$ belong to the same
coset, then
\begin{equation}
  O\mathbf{e}_1 = O'\mathbf{e}_1 = \mathbf{\hat v}
\end{equation}
and vice versa.  In other words, cosets are specified by where
$\mathbf{e}_1$ is mapped. Furthermore, since $\norm{O\mathbf{e}_1}=1$,
we see that they can be identified with the points in the unit sphere.
Finally, the map~\eqref{H_rm} is one-to-one and is such that
\begin{equation}
  H_N(\mathbf{\hat v})\mathbf{e}_1=\mathbf{\hat v}.
\end{equation}
Therefore, $H_N$ spans a complete class of representatives.
\end{proof}

Incidentally, theorem~\ref{hpr} implies
\begin{corollary}
  \label{fac_mc}
  Let $d\mu_{\ON}$ and $d\mu_{\mathrm{O}(N-1)}$ be the Haar measures
  on $\ON$ and $\mathrm{O}(N-1)$ respectively.  Then
  \begin{equation}
    \label{fact_m}
    d\mu_{\ON} =  d\mu_{\mathbb{S}^{N-1}}\times
    d\mu_{\mathrm{O}(N-1)},
  \end{equation}
  where $d\mu_{\mathbb{S}^{N-1}}$ is the uniform measure on
  $\mathbb{S}^{N-1}$.
\end{corollary}
What is the meaning of $d\mu_{\mathbb{S}^{N-1}}$? Given that we are
dealing with uniform random variables, it is quite natural that we end
up with the uniform measure.  In this case, however, it has a precise
group theoretical interpretation. Left multiplication of the
right-hand side of equation~\eqref{facmat} by $O' \in \ON$ induces a
map on the coset space:
\begin{equation}
  O'H_N(\mathbf{\hat v}) \Omega = H_N(\mathbf{\hat v}')\Omega'
  \Omega = H_N(\mathbf{\hat v}')\Omega'' .
\end{equation}
Since the decomposition~\eqref{facmat} is unique the transformation
$\mathbf{\hat v} \mapsto \mathbf{\hat v}'$ is well defined.  This map
can be easily determined.  A coset is specified by where
$\mathbf{e}_1$ is mapped, therefore
\begin{equation}
  O'H_N(\mathbf{\hat v})\mathbf{e}_1 = O'\mathbf{\hat v} =
  \mathbf{\hat v}' = H_N(\mathbf{\hat v'})\mathbf{e}_1.
\end{equation}
If $\mathbf{\hat v}$ is uniformly distributed on the unit circle
so is $\mathbf{\hat v}'=O\mathbf{\hat v}$.  Thus,
$d\mu_{\mathbb{S}^{N-1}}$ is the unique measure on the coset space
$\ON/\mathrm{O}(N-1)$ invariant under the left action of $\ON$.
Its uniqueness follows from that of Haar measure and from the
factorization~\eqref{fact_m}.

Corollary~\ref{fac_mc} is a particular case of a theorem that
holds under general hypotheses for topological compact groups.
Indeed, let $\Gamma$ be such a group, $\Xi$ a closed subgroup and
$C= \Gamma/\Xi$.  Furthermore, let $d\mu_{\Gamma}$, $d\mu_{C}$ and
$d\mu_{X}$ be the respective invariant measures, then
\begin{equation}
  \label{gmesfact}
  d\mu_{\Gamma} =  d\mu_{\Xi} \times d\mu_{C}.
\end{equation}

%\section{The COE and CSE ensembles}
%\label{circular}

%These ensembles play a very important role in Physics to model
%quantum dynamical systems.  In this section, we briefly discuss
%how to generate such random matrices.

%The space of matrices in the COE is a complete set of
%representatives of the quotient space $\UN/\ON$; the probability
%distribution is the unique invariant measure induced on $\UN/\ON$
%by Haar measure on $\UN$.  A class of representatives is
%constructed using the representation
%\begin{equation}
%S=UU^t,
%\end{equation}
%for any $U \in \UN$. (See Mehta~\cite{Meh04} pp.~192--194.) From
%the decomposition~\eqref{gmesfact}, it follows that $S$ is in the
%COE if $U$ is distributed according to Haar measure on $\UN$.

%The CSE is analogous (Mehta~\cite{Meh04} pp.~194--196).  It is
%defined as the quotient space \newline $\mathrm{U}(2N)/\USpN$ with
%the corresponding invariant measure. A class of representatives is
%constructed as follows. Take $U\in \mathrm{U}(2N)$; it has a
%representation in terms of an $N\times N$ quaternion matrix
%$\mathcal{U}$. (In this case, however, the coefficients of the
%quaternion units are allowed to be complex.) Consider the set of
%quaternion matrices
%\begin{equation}
%\mathcal{W} = \mathcal{U}\mathcal{U}^*.
%\end{equation}
%From equation~\eqref{dual} $\mathcal{W}$ is mapped into
%\begin{equation}
%W= -UJU^tJ
%\end{equation}
%and span a complete set of representatives of
%$\mathrm{U}(2N)/\USpN$. If $U$ is distributed with Haar measure on
%$\mathrm{U}(2N)$, then $W$ is uniformly distributed in the CSE.

\section*{Acknowledgements}
This article stems from a lecture that I gave at the summer school on
Number Theory and Random Matrix Theory held at the University of
Rochester in June 2006.  I would like to the thank the organisers
David Farmer, Steve Gonek and Chris Hughes for inviting me. I am also
particularly grateful to Brian Conrey, David Farmer and Chris Hughes
for the encouragement to write up the content of this lecture.

\vspace{.25cm}

\noindent\rule{16.2cm}{.5pt}

\vspace{.25cm}

{\small \noindent {\sl School of Mathematics \\
                       University of Bristol\\
                       Bristol BS8 1TW, UK  \\
                       Email: {\tt f.mezzadri@bristol.ac.uk}

                       \vspace{.25cm}

                       \noindent 27 February 2007}}

\end{document}